\documentclass[conference,twocolumn]{IEEEtran}
\usepackage{amsmath, amssymb, amsfonts, latexsym}
\usepackage{graphicx}
\newtheorem{theorem}{Theorem}
\newtheorem{lemma}{Lemma}
\newtheorem{definition}{Definition}

\def\FF{\mathbb{F}}

\newcommand{\tran}[1]{#1^T}

\DeclareMathOperator{\head}{head}
\DeclareMathOperator{\tail}{tail}

\newcommand{\edmonds}{Z}
\newcommand{\graph}{\mathcal{G}}
\newcommand{\edge}{e}
\newcommand{\node}{v}
\newcommand{\edges}{\mathcal{\MakeUppercase{\edge}}}
\newcommand{\nodes}{\mathcal{\MakeUppercase{\node}}}
\newcommand{\numedges}{\MakeUppercase{\edge}}
\newcommand{\numnodes}{\MakeUppercase{\node}}
\newcommand{\din}[1]{d_{\text{in}}\left(#1\right)}
\newcommand{\dout}[1]{d_{\text{out}}\left(#1\right)}

\newcommand{\event}{\mathsf{E}}

\title{On Random Network Coding for Multicast}
\author{
\authorblockN{Adrian Tauste Campo}
\authorblockA{Universitat Pompeu Fabra\\Barcelona, Spain}
\and
\authorblockN{Alex Grant}
\authorblockA{Institute for Telecommunications Research\\
  University of South Australia}
}

\begin{document}
\maketitle

\begin{abstract}
  Random linear network coding is a particularly decentralized
  approach to the multicast problem. Use of random network codes
  introduces a non-zero probability however that some sinks will not
  be able to successfully decode the required sources. One of the main
  theoretical motivations for random network codes stems from the
  lower bound on the probability of successful decoding reported by Ho
  et. al. (2003). This result demonstrates that all sinks in a
  linearly solvable network can successfully decode all sources
  provided that the random code field size is large enough. This paper
  develops a new bound on the probability of successful decoding.
\end{abstract}

\section{Introduction}
It has been recently proved that network layer coding can increase
throughput, particularly for multicast scenarios \cite{AhlCai00}. It
is also known that linear network codes \cite{LiYeu03} can achieve
max-flow upper bounds on the throughput in a single source multicast
network. The algebraic approach of \cite{KoeMed03} is particularly
useful in the design and analysis of linear network codes, and we
adopt the notation and terminology of that paper. 

Random networks codes \cite{HoMed03,HoKoe03} are linear network codes
in which the encoding coefficients are chosen randomly from a finite
field. The sink nodes can decode correctly if and only if the overall
transfer matrix from the sources to each sink is invertible. One of
the main theoretical results for random network codes consists of the
following lower bound on the probability of successful decoding
\cite{HoMed03}, assuming that the underlying network is linearly
solvable over $\FF_q$ (i.e. there exists a linear code which satisfies
the multicast requirements).  For a network code in which some of the
code coefficients are chosen independently and uniformly from a finite
field with cardinality $q$, the probability that all $d$ receivers can
decode the source processes is at least
\begin{equation}\label{eq:bound}
  \left(1-\frac{d}{q}\right)^{\nu}
\end{equation}
where $\nu$ is the maximum number of links receiving signals with
independent random coefficients in any set of links constituting a
flow solution from all sources to any receiver \cite{HoKoe03}. 

A looser bound (subject to the same conditions as above) which depends
only on $\eta$, the total number of edges receiving signals with
independent random coefficients is given by \cite{HoMed03,HoMed}
\begin{equation}
  \label{eq:bound2}
  \left(1-\frac{d}{q}\right)^\eta.
\end{equation}

Thus provided a linear solution over $\FF_q$ exists in the first
place, the probability of successful decoding can be made as close to
one as desired, by increasing the field size $q$. The bounds
(\ref{eq:bound}) and (\ref{eq:bound2}) rely on the special structure
of the determinant polynomial of the transfer matrix of the network.

This paper develops the following new lower bound.

\begin{theorem}
  Consider a network code in which $\eta$ edges receive signals with
  independent random coefficients chosen independently and uniformly
  from a finite field with cardinality $q$.  If there is some choice
  of coefficients for these $\eta$ edges that results in a solution
  over $\FF_q$ then the probability that all receivers can decode the
  source processes is at least
  \begin{equation}
    \label{eq:newbound}
    \left(1-\frac{1}{q}\right)^{\eta}.
  \end{equation}
\end{theorem}

Our approach for the proof of this theorem is to identify a critical
sub-matrix of the Edmonds matrix whose non-singularity is a necessary
and sufficient condition for decoding success. This critical matrix is
different for each sink in the network.  The new bound results
directly from a nesting property of the critical matrices.

In the new bound, the field size $q$ required to attain a given
probability of success depends only on the number of edges with random
coefficients, and not on the number of sinks. The resulting $d$-fold
reduction in the required $q$ could be significant. We emphasize that
(\ref{eq:newbound}), like (\ref{eq:bound}) applies only when the
underlying network is solvable over $\FF_q$. This is a consequence of
the conditions for applicability of the Schwartz-Zippel inequality,
which is used in the proof of both bounds. Thus (\ref{eq:newbound})
does \emph{not} imply the universal existence of binary solutions for
every network. The bounds (\ref{eq:bound}),  (\ref{eq:bound2}) and
(\ref{eq:newbound}) only provide lower bounds for a given $q$ when the
network is solvable over $\FF_q$.

We further conjecture that for large random networks satisfying
certain properties, the success probability behaves as
\begin{equation}
  \label{eq:conjecture}
  \prod_{i=1}^{\numedges} \left(1-\frac{1}{q^i}\right)
\end{equation}
where $\numedges$ is the total number of links in the network.

The paper is organized as follows: Section \ref{sec:model} presents
our model and introduces some algebraic notation.  Section
\ref{sec:newbound} develops the new bound (\ref{eq:newbound}), while
Section \ref{sec:random} discusses random graphs, leading to the
conjecture (\ref{eq:conjecture}).

\section{Network Coding Model}\label{sec:model}
We adopt the model from \cite{KoeMed03}. The network is represented by
a directed acyclic graph $\graph=(\nodes,\edges)$ with
$\numnodes=|\nodes|$ nodes and $\numedges=|\edges|$ edges. There are
$r$ independent, discrete source processes with messages
belonging to $\FF_q$, and $d\geq 1$ receivers. Each receiver
node has $L\geq r$ incoming edges. The multicast requirement is that
each receiver node can decode every source message from the signals on
its incident edges.

Each edge $\edge\in\edges$ is incident to node $\node\in\nodes$ if
$\node=\head(\edge)$, or is an outgoing edge if $\node=\tail(l)$. The
in-degree of a node $\node$ is $\din{\node}$ and the out-degree is
$\dout{\node}$. The time unit is chosen such that the capacity of each
link is one bit per unit time and edges with larger capacity are
modeled as parallel edges.  Without loss of generality, it can be
assumed that each source is associated with a source node
$s_\alpha\in\nodes$ with $\din{s_\alpha}=0$ and $\dout{s_\alpha}=1$,
$\alpha=1,2,\dots,r$ . Similarly, each sink node $t_\beta$ has
$\din{t_\beta}=r$ and $\dout{t_\beta}=0$, $\beta=1,2,\dots,d$ (it is
always possible to obtain such a graph by introducing auxiliary nodes
and edges). It will further be assumed that edges are labeled
ancestrally.

A \emph{scalar linear network code} for $G$ is an assignment of linear
encoding functions $f_v:\FF_q^{\din{v}}\mapsto\FF_q^{\dout{v}}$ to
each node $v\in\nodes$. Such codes are sufficient for the multicast
problem on acyclic delay networks.  Following \cite{KoeMed03}, define
the \emph{encoding matrix} $F\in\FF_q^{\numedges\times\numedges}$
where $F_{ij}$ is the coefficient applied to the symbol incoming on
edge $i\in\edges$ for contribution to outgoing edge
$j\in\edges$. According to the assumption of ancestral ordering, $F$
is strictly upper triangular.  Similarly, the \emph{source matrix}
$A\in\FF_q^{r\times\numedges}$ maps messages onto outgoing source
edges and the \emph{sink matrix}
$B_{\beta}\in\FF_q^{r\times\numedges}$ maps incoming sink edges onto
the sinks $t_\beta\in\nodes$, $\beta=1, 2,\dots, d$.
% Further define
% \begin{equation*}
% B =
% \begin{bmatrix}
%   B_1       \\
%   B_2
%   \vdots    \\
%   B_d
% \end{bmatrix}
% \end{equation*}

Let $x\in\FF_q^{1\times r}$ be a row vector representing the source
messages. Then the received vector of symbols
$y_\beta\in\FF_q^{1\times r}$ at sink 
$\beta=1,2,\dots,d$ is given by
\begin{displaymath}
  y_\beta = x M_\beta
\end{displaymath}
where 
\begin{equation*}
  M_\beta = A (I-F)^{-1} \tran{B_\beta}.  
\end{equation*}
Each sink can decode all sources if and only if $\det(A (I-F)^{-1}
\tran{B_\beta}) \neq 0$ for every $\beta=1,2,\dots,d$, or equivalently
if the Edmonds matrix
\begin{equation*}
  \edmonds_\beta=
  \begin{bmatrix}
    A & 0 \\
    I-F & \tran{B_\beta}
  \end{bmatrix}
\end{equation*}
is non-singular. 

Considering the entries of $A$, $F$ and $B_\beta$ as variables, the
Leibniz determinant formula provides a way of writing $\det
\edmonds_\beta$ as a multivariate polynomial $P_\beta$ in the $a_{ij},
f_{ij}, b_{ij}$. Furthermore, this multivariate polynomial has degree
at most $\nu$ but is linear in each variable individually. Therefore
the product
\begin{equation}
  P=\prod_\beta P_\beta\label{eq:P}
\end{equation}
has degree $d\nu$, with each variable of degree $d$ or less.

The lower bound (\ref{eq:bound}) results from a modified
Schwartz-Zippel bound, which takes into account the individual
variable degree constraint of $P_\beta$ \cite[Lemma 1]{HoKoe03}. We
reproduce this lemma for reference.
\begin{lemma}\label{lem:sz}
  Let $P$ be a multivariate polynomial of degree $d\nu$, with the
  exponent of any individual variable at most $d$. Let each variable
  be chosen uniformly from $\FF_q$. Then if $P$ is not identically
  zero,
  \begin{equation}\label{eq:sz}
    \Pr\left(P\neq0\right) \geq \left(1-\frac{d}{q}\right)^\nu.
  \end{equation}
\end{lemma}

We make two remarks on this approach. First, application of Lemma
\ref{lem:sz} to $P$ as defined in (\ref{eq:P}) implies an independence
of the events $P_{\beta_1}=0$ and $P_{\beta_2}=0$. Depending on the
structure of the network, these events may be strongly dependent. For
example, consider $P_{1}=P_{2}=\dots=P_d$, meaning all sinks have
identical incoming signals ($B_1=B_2=\dots=B_d$). Then Lemma
\ref{lem:sz} yields a lower bound $(1-d/q)^\nu$, rather than
$(1-1/q)^\nu$. Obviously this is an extreme example, yet it
illustrates the point that (\ref{eq:bound}) may be loose.

Secondly, the modified Schwartz-Zippel bound itself can be very loose,
as the following example shows. Let $H\in\FF_q^{m\times m}$ with each
entry $h_{ij}$ chosen independently with a uniform distribution on
$\FF_q$. Then it is well known that
\begin{equation}
  \label{eq:fullexact}
  \Pr\left(\det H \neq 0\right) = \pi_m(q) = \prod_{i=1}^m
  \left(1-q^{-i}\right). 
\end{equation}
In contrast, Lemma \ref{lem:sz} gives the lower bound
\begin{equation}\label{eq:badbound}
  \Pr\left(\det H \neq 0\right) \geq \left(1-q^{-1}\right)^m, 
\end{equation}
which also could be obtained from (\ref{eq:fullexact}) by lower
bounding each term in the product by the minimum term
$(1-q^{-1})$. 

We emphasize that (\ref{eq:sz}) applies only when $P$ is not
identically zero for every choice of variables (e.g. all coefficients
are zero). This precludes application of (\ref{eq:sz}) to non-solvable
networks, i.e. networks where every choice of $F$ makes $Z_\beta$
singular and hence $P=0$.

In Section \ref{sec:newbound} we partially address the dependency
between the $P_\beta$, while in Section \ref{sec:random} we consider
large random networks, where we also discuss the extent to which
(\ref{eq:fullexact}) improves (\ref{eq:badbound}).

\section{The New Bound} \label{sec:newbound} According to our
assumption regarding sources and sinks, and the ancestral ordering of edges,
we can further assume without loss of generality that
\begin{align*}
  A &=
  \begin{bmatrix}
    I_{r\times r} & 0_{r\times (\numedges-r)}
  \end{bmatrix}
  \\
  B_\beta &=
  \begin{bmatrix}
    0_{r\times k_\beta} & I_{r\times r} & 0_{r\times(\numedges-r-k_\beta)}
  \end{bmatrix}, \beta=1,2,\dots,d
\end{align*}
where $k_1>r$ and $k_\beta > r+k_{\beta-1}$, $\beta>1$. This means
that the sources inject messages into the network via edges
$1,2,\dots,r$ and that each sink observes signals on $r$ consecutively
numbered edges. No sink shares edges with any other sink or
source. See Figure \ref{fig:butterfly} for an example of how to
arrive at this formulation.

Then the Edmonds matrix for sink $\beta$ has the following structure:
\begin{equation}
  Z_\beta =
  \begin{bmatrix}
    I_{r} & 0   & 0   & 0 & 0      \\
    U_1           & W_{11} & W_{12} & W_{13} & 0 \\
    0             & U_2 & W_{21} & W_{22} & 0 \\
    0             & 0   & U_3     & W_{31} & I_{r} \\
    0             & 0   & 0       & U_4   & 0
  \end{bmatrix}
\end{equation}
where the $U_i$ are square, upper triangular with diagonal elements all
equal to $1$. The matrices $U_1$ and $U_3$ are  $r\times r$, $U_2$ is
$(k_\beta-2r)\times(k_\beta-2r)$ and $U_4$ is
$(\numedges-r-k_\beta)\times(\numedges-r-k_\beta)$. 

\begin{definition}
  The \emph{critical matrix} for sink $\beta$ is the following
  $(k_\beta-r)\times(k_\beta-r)$ principal sub-matrix of $Z_\beta$,
  \begin{equation}\label{eq:critical}
    C_\beta = \begin{pmatrix}
      W_{11} & W_{12} \\
      U_2 & W_{21}
    \end{pmatrix}.
  \end{equation}
\end{definition}

\begin{lemma}\label{lem:critical}
The determinant of the Edmonds matrix for sink $\beta$ has the same
magnitude as the determinant of its critical matrix. 
\begin{equation*}
  |\det Z_\beta| = |\det C_\beta| 
\end{equation*}
\end{lemma}
\begin{proof}
  Straightforward from either the Laplace expansion of $\det Z_\beta$,
  or repeated application of the partitioned matrix determinant formula.
\end{proof}
We can immediately apply Lemma \ref{lem:sz} to $\det C_\beta$ to bound
the probability for a given sink
\begin{equation}\label{eq:onesink}
  \Pr\left(\det Z_\beta\neq 0\right) = \Pr\left(\det C_\beta\neq
    0\right) \geq \left(1-\frac{1}{q}\right)^{\eta_\beta},
\end{equation}
where $\eta_\beta$ is the number of columns in $C_\beta$ with variable
terms, i.e. the number of edges in the subset $\{r+1,r+2,\dots,k_\beta\}$
receiving signals with random coefficients.

For the $d$ receiver problem, we have the following very useful
property of the critical matrices, which is guaranteed by their
construction.
\begin{lemma}[Nesting of critical matrices]\label{lem:nesting}
  $C_{\beta_1}$ is a principal
  sub-matrix of $C_{\beta_2}$ for $\beta_2 > \beta_1$.  
\end{lemma}
Hence each critical matrix $C_\beta$ has as nested principal
sub-matrices, all the critical matrices for sinks $1,2,\dots,\beta-1$.

\begin{proof}[Proof of main result (\ref{eq:newbound})]
  Let $\event_\beta$, $\beta=1,2,\dots,d$ be the event that sink $\beta$ can
  decode. By Lemma \ref{lem:critical}, $\event_\beta \iff
  \det Z_\beta \neq 0 \iff \det C_\beta \neq 0$. Now the probability
  that all sinks can decode is given by
  \begin{equation}\label{eq:chainrule}
    \Pr\left(\bigcap_{\beta=1}^d \event_\beta\right) =
    \Pr(\event_1) \Pr(\event_2\mid\event_1) \dots
    \Pr(\event_\beta\mid\event_1\dots\event_{\beta-1})   
  \end{equation}
  Now consider $\Pr(\event_m\mid\event_1,\dots,\event_{m-1}) = \Pr(\det
  C_m\neq 0 \mid \det C_1 \neq 0,\dots,\det C_{m-1}\neq0)$ for some $2\leq
  m\leq\beta$. By Lemma \ref{lem:nesting}, $C_m$ can be partitioned
  \begin{equation*}
    C_m =
    \begin{pmatrix}
      C_{m-1} & U \\
      V & W
    \end{pmatrix}
  \end{equation*}
  for appropriate choices of $U, V, W$. 

  Conditioned on $\det C_{m-1} \neq 0$, we can use the partitioned
  matrix determinant formula to write
  \begin{equation}\label{eq:partitiondet}
    \det C_m = \det(C_{m-1}) \det\left(W - V C_{m-1}^{-1} U\right),
  \end{equation}
  which (conditioned on $\det C_{m-1} \neq 0$) is zero if and only if
  $\det\left(W - V C_{m-1}^{-1} U\right)=0$.

  Let $\phi_m$ be the multivariate polynomial corresponding to $\det
  C_m$, and let $\sigma_{m-1}$ be the multivariate polynomial
  corresponding to $\det\left(W - V C_{m-1}^{-1} U\right)$. Then from
  (\ref{eq:partitiondet}) $\deg\phi_m = \deg\phi_{m-1} +
  \deg\sigma_{m-1}$. This relation also holds for the degree of any
  individual variable. From the Leibniz formula and the structure of
  the Edmonds matrix (as explained previously for $P_\beta$), we also
  know that the individual degree of any variable in $\phi_m$ or
  $\phi_{m-1}$ is zero or one.  Hence
  \begin{equation*}
    \deg\sigma_{m-1} = \deg\phi_m - \deg\phi_{m-1},
  \end{equation*}
  and the degree of any individual variable in $\sigma_{m-1}$ is at
  most 1.  Collecting results so far and applying Lemma \ref{lem:sz},
  \begin{align*}
    \Pr(\event_m\mid\event_1,\dots,\event_{m-1})
    &= \Pr\left(\det\left(W - V C_{m-1}^{-1} U\right)\neq 0 \right) \\
      &=
    \Pr\left(\sigma_{m-1}\neq 0\right) \\ &\leq
    \left(1-\frac{1}{q}\right)^{\deg\phi_m - \deg\phi_{m-1}} 
  \end{align*}
  
  Finally, substitution into (\ref{eq:chainrule}) results in a
  telescoping sum for the exponents, $\deg\phi_1 + \deg\phi_2 -
  \deg\phi_1 + \deg\phi_3 - \deg\phi_2 + \dots$, leaving only
  \begin{equation*}
    \Pr\left(\bigcap_{\beta=1}^d \event_\beta\right) \geq \left(1 -
      \frac{1}{q}\right)^{\deg\phi_d} 
  \end{equation*}

  This directly yields (\ref{eq:newbound}) via $d\nu \leq \eta
  \triangleq \deg\phi_d = \eta_d \leq \numedges$.
\end{proof}

Let 
\begin{equation*}
  z(d,q)=\frac{\log(1-d/q)}{\log(1-1/q)}.
\end{equation*}
Then (\ref{eq:newbound}) is tighter than (\ref{eq:bound}) whenever
\begin{equation*}
  \eta < \nu\, z(d,q).
\end{equation*}
Furthermore, $z(d,q)>d$ and
\begin{align*}
  \lim_{q\rightarrow d}z(d,q)&=\infty \\
  \lim_{q\rightarrow\infty}z(d,q)&=d.
\end{align*}
Roughly speaking, the new bound is tighter for networks with
$\numedges = O(\nu d)$ and sufficiently small $q$.

In some instances it may be useful to have a bound which depends only
on the total number of edges carrying signals with random
coefficients. Replacing $\nu$ with $\eta$ in (\ref{eq:bound}) results
in (\ref{eq:bound2}) which is looser than (\ref{eq:newbound}), since
\begin{equation*}
   \left(1-d/q\right)^\eta < \left(1-1/q\right)^\eta.
\end{equation*}

Note that successful decoding at a particular sink $\beta$ in general
depends on only part of $C_\beta$. There can be a much smaller
sub-matrix that determines singularity, for example, $C_\beta$ might
be block diagonal, with successful decoding of sink $\beta$ depending
only on one of the blocks (this case arises when there are disjoint
paths from the sources to each sink). Thus $C_\beta$ may be larger
than strictly required for analysis of sink $\beta$ alone, however
defining the critical matrix this way yields the nesting property that
results in the new bound.

\section{Example: The Butterfly Network}
\begin{figure}[htbp]
  \centering
  \includegraphics*[scale=0.7]{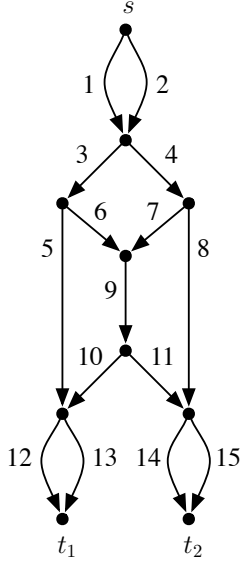}
  \caption{The butterfly network.}
  \label{fig:butterfly}
\end{figure}

Figure \ref{fig:butterfly} shows the well-known butterfly network,
with additional nodes and edges introduced in order to satisfy our
assumptions on sources and sinks. The source $s$ has $r=2$ messages,
and the edge labels indicate the edge ordering. Edges $1$ and $2$
carry the two messages from the source, while edges $12$ resp. $13$
duplicate the signals on edges $5$ resp. $10$, and edges $14$
resp. $15$ duplicate $8$ resp. $11$. Supposing that all other edges
carry random linear combinations of signals, $\nu=7$ and $\eta = 9$.

\begin{figure}[htbp]
  \centering
  \includegraphics*[width=0.8\columnwidth]{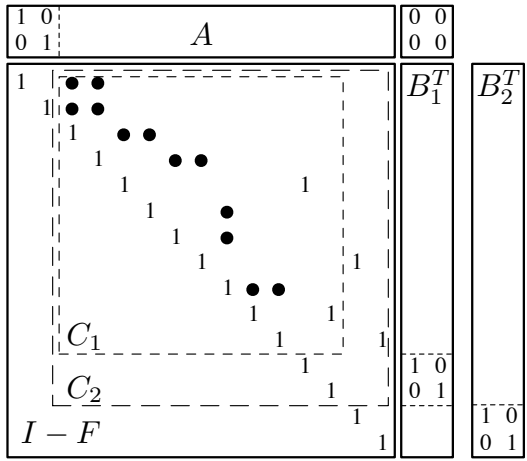}
  \caption{Critical matrices for the butterfly network.}
  \label{fig:critical}
\end{figure}

Figure \ref{fig:critical} shows the structure of the Edmonds matrix
$Z_1$, and the nested critical matrices $C_1$ and $C_2$. To see how
the nesting arises, $B_2$ has been placed alongside. For clarity, most
of the zeros have been omitted from each matrix. The solid disks
represent random entries of $F$.

\begin{figure}[htbp]
  \centering
  \setlength{\unitlength}{1mm}
  {\begin{picture}(85,55)
    \put(0,0){\includegraphics*[width=0.9\columnwidth]{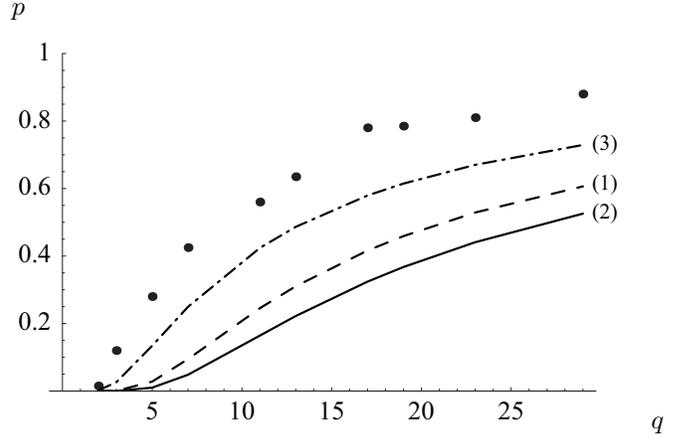}}
    \put(85,0){\makebox(0,0){$q$}}
    \put(0,55){\makebox(0,0){$p$}}
  \end{picture}}
  \caption{Success probability $p$ versus field size $q$ compared to
    bounds (\ref{eq:bound}), (\ref{eq:bound2}) and
    (\ref{eq:newbound}) for the butterfly network.}
  \label{fig:plot}
\end{figure}

Figure \ref{fig:plot} shows the empirically measured probability of
decoding success versus the field size $q$ for the network of Figure
\ref{fig:butterfly} (filled circles). This was achieved using monte-carlo
simulation, selecting each of the coefficients uniformly from $\FF_q$.
Results for the first ten prime fields are shown. Also shown are the
existing bounds (\ref{eq:bound}), dashed line, (\ref{eq:bound2}), solid
line, and the new bound (\ref{eq:newbound}), dot-dashed line. In this
case, the new bound is considerably tighter.

%  The matrices $A$,
% $B_1$, $B_2$ and $F$ have the following form.
% \begin{align*}
%   A &=
%   \begin{bmatrix}
%     I_2 & 0_{13}
%   \end{bmatrix} \\
%   B_1 &=
%   \begin{bmatrix}
%     0_{11} & I_2 & 0_{2}
%   \end{bmatrix} \\
%   B_2 &=
%   \begin{bmatrix}
%     0_{13} & I_2 
%   \end{bmatrix} \\
% I-F &=
% \left[
% \begin{array}{ccccccccccccccc}
%  1&&x&x&&&&&&&&&&& \\
%  &1&x&x&&&&&&&&&&& \\
%  &&1&&x&x&&&&&&&&& \\
%  &&&1&&&x&x&&&&&&& \\
%  &&&&1&&&&&&&x&&& \\
%  &&&&&1&&&x&&&&&& \\
%  &&&&&&1&&x&&&&&& \\
%  &&&&&&&1&&&&&&&x \\
%  &&&&&&&&1&x&x&&&& \\
%  &&&&&&&&&1&&&x&& \\
%  &&&&&&&&&&1&&&x& \\
%  &&&&&&&&&&&1&&& \\
%  &&&&&&&&&&&&1&& \\
%  &&&&&&&&&&&&&1& \\
% \end{array}\right]
% \end{align*}

\section{Random Graphs}\label{sec:random}
Successful decoding for a particular sink $\beta$ depends on the
non-singularity of its critical matrix $C_\beta$. To obtain
(\ref{eq:newbound}) we used Lemma \ref{lem:sz} to bound the
probability that this matrix is non-singular. It is interesting to
consider however circumstances under which (\ref{eq:fullexact}) might
be applicable, providing an even tighter bound.

There are two main obstacles to the application of
(\ref{eq:fullexact}) for determination of the probability that $\det
C_\beta\neq 0$. Firstly, (\ref{eq:fullexact}) applies to ``full''
matrices, with each element chosen independently and uniformly from
$\FF_q$.  In contrast, $C_\beta$ is of the form (\ref{eq:critical}),
with all elements below the $r$-th diagonal equal to zero (the
strictly lower triangular part of $U_2$). Secondly, the non-zero
elements in the upper portion (upper triangular part of $U_2$ and all
of $W_{11}$, $W_{12}$ and $W_{21}$) of $C_\beta$ are determined by the
topology of the network itself. For a sparsely connected network, the
proportion of zeros in this part of the matrix will greatly exceed
$1/q$.

Assuming that the random network code coefficients are chosen from the
non-zero elements of $\FF_q$, the total number of non-zero elements in
$F$ is
\begin{equation*}
  \sigma \triangleq \sum_{\node\in\nodes} \din{\node}\dout{\node} \leq
  \numedges^2. 
\end{equation*}
Let $\rho=\sigma/E^2$ be the proportion of non-zero elements. Ignoring
the structure required by (\ref{eq:critical}), generate a random
$m\times m$ matrix $C^{(m)}$ with elements identically distributed
according to
\begin{equation*}
  \Pr\left(c_{ij}=f\right) =
  \begin{cases}
    1-\rho & f=0 \\
    \frac{\rho}{q-1} & f \neq 0
  \end{cases}
\end{equation*}

It is a remarkable fact that provided $\rho$ does not tend to zero or one
too quickly with $m$, 
\begin{equation*}
  \lim_{m\rightarrow\infty} \Pr\left(\det C^{(m)}\neq 0\right) = \pi_m(q).
\end{equation*}
See \cite{Coo00} for a discussion of this threshold
effect. Conditioned on the event that $C^{(m)}$ has no all-zero rows
or columns (if it did, the network flow would anyway be infeasible
regardless of choice of code), the requirement is
\begin{equation*}
  \rho > \frac{1}{m}\left(\frac{1}{2}\log m + \log\log m\right).
\end{equation*}
This result even holds for independent, but non-identically
distributed entries, as discussed by Cooper \cite{Coo00}.  

Now for sufficiently small $\rho$, $C^{(m)}$ can be permuted with high
probability into the form (\ref{eq:critical}). This leads us to
conjecture that there exist conditions on $\sigma$ such that
$\pi_m(q)$ is the success probability for a large, randomly generated
network with a given degree distribution. The remainder of this
section analyzes some properties of $\pi_m(q)$, and demonstrates the
improvement that may be obtained compared to (\ref{eq:badbound}).

To guarantee a particular probability $p$ using (\ref{eq:badbound}),
the field size $q$ must satisfy
\begin{equation*}
  q \geq \frac{1}{1-p^{1/m}} =  \frac{1}{2}+ m \log \frac{1}{p} +
  O\left(\frac{1}{m}\right).  
\end{equation*}
Hence the required field size increases linearly with the size of the
matrix.

% \begin{equation*}
%   \lim_{m\rightarrow\infty}\left(1-1/q\right)^m = 0
% \end{equation*}

\begin{figure}
  \centering
  \setlength{\unitlength}{1mm}
  {\begin{picture}(80,55)
    \put(0,0){\includegraphics*[width=80mm]{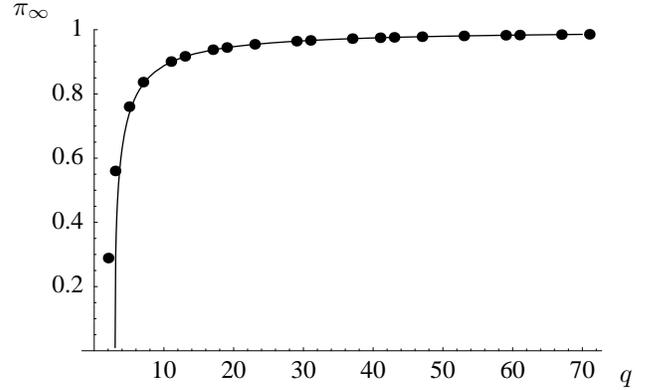}}
    \put(80,1){\makebox(0,0){$q$}}
    \put(1,50){\makebox(0,0){$\pi_\infty$}}
  \end{picture}}
  \caption{Lower bound (solid line) and $\pi_\infty(q)$ (dots).}
  \label{fig:bound}
\end{figure}

Let $\pi_\infty(q)=\lim_{m\rightarrow\infty}\pi_m(q)$ then
\begin{equation*}
  \pi_\infty(q) = \prod_{i=1}^\infty\left(1-q^{-i}\right) = q^{1/24}
  \left(\frac{1}{2} 
  \vartheta_1'\left(q^{-1/2}\right)\right)^{1/3},
\end{equation*}
where $\vartheta_1$ is the Jacobi theta function \cite[Equation
8.181.3]{GraRyz94} and
\begin{align*}
  \vartheta_1'(q) &= \left.\frac{\partial}{\partial
      z}\vartheta_1(z,q)\right|_{z=0} \\
  &= 2\sum_{i=0}^\infty (-1)^i (1+2i) q^{-\frac{1}{2}(i+\frac{1}{2})^2}.
\end{align*}
Truncating the latter series gives the following lower bound,
\begin{equation*}
  \pi_\infty(q) \geq \left(1-\frac{3}{x}\right)^{1/3}.
\end{equation*}
This lower bound is compared to $\pi_\infty$ for the first 20 primes
in Figure \ref{fig:bound}. 
For a given probability $p$ in (\ref{eq:fullexact}), the required
field size $q$ for $m\rightarrow\infty$ satisfies
\begin{equation*}
  q \geq \frac{3}{1-p^3}.
\end{equation*}
which does not depend on $m$.

\section{Concluding remarks}
\label{sec:conclusion}
Random network coding is a promising decentralized approach for
multicast. One of the main implementation considerations is the size
of the finite field required to achieve a specified probability that
every sink can decode every source. This paper presented a new bound
on the success probability, which in certain circumstances is tighter
that the previous bound. We also presented a heuristic argument that
motivates the investigation of tighter bounds for large random
networks, based on the distribution of rank of large random finite
field matrices. 

\section*{Acknowledgments}
This work was performed while A. Tauste Campo was visiting the
Institute for Telecommunications Research.  This work was supported by
the Australian Government under grant DP0557310, and by the Defence
Science and Technology Organisation under contracts 4500485167 and
4500550654. The authors would like to thank Ian Grivell and Terence
Chan and for helpful discussions.

\bibliographystyle{IEEEtran}
\bibliography{network,alex}

\end{document}